\documentclass[useAMS,usenatbib]{mn2e}

\def\strutdepth{\dp\strutbox}
\def\marginalstar{\strut\vadjust{\kern-\strutdepth\specialstar}}
\def\specialstar{\vtop to \strutdepth{\baselineskip\strutdepth\vss\llap{$\bigtriangleup\mskip-12mu\bigtriangledown$ }\null}}
\def \nm {N_{\mathrm{m}}}
\def \nm0 {N_{\mathrm{m,0}}}
\def \na {N_{\mathrm{a}}}
\def \na0 {N_{\mathrm{a,0}}}

\def \BGE {\begin{equation}}
\def \EDE {\end{equation}}

\title[Night Sky brightness at sites from DMSP-OLS satellite measurements]{Night sky brightness at sites from DMSP-OLS satellite measurements}

   \author[P. Cinzano and C. D. Elvidge]{P. Cinzano$^{1,2}$\thanks{E-mail: cinzano@pd.astro.it}
          and C. D. Elvidge$^3$ \\
$^{1}$ Dipartimento di Astronomia, Universit\`a di Padova,
Vicolo dell'Osservatorio 2,  I-35122 Padova, Italy\\
$^2$ Istituto di Scienza e Tecnologia dell'Inquinamento Luminoso
(ISTIL), Via Roma 13, I-36016 Thiene, Italy\\
 $^3$ NOAA National
Geophysical Data Center, 325 Broadway, Boulder CO 80303}

\date{Accepted 17 June 2004.
      Received 26 May 2004;
      in original form 20 April 2004}

\pagerange{\pageref{firstpage}--\pageref{lastpage}} \pubyear{2004}

\input epsf.sty
\begin{document}

\maketitle

\label{firstpage}

\begin{abstract}

We apply the sky brightness modelling technique introduced and
developed by Roy Garstang to high-resolution DMSP-OLS satellite
measurements of upward artificial light flux and to GTOPO30
digital elevation data in order to predict the brightness
distribution of the night sky at a given site in the primary
astronomical photometric bands for a range of atmospheric aerosol
contents. This method, based on global data and accounting for
elevation, Earth curvature and mountain screening, allows the
evaluation of sky glow conditions over the entire sky for any site
in the World, to evaluate its evolution, to disentangle the
contribution of individual sources in the surrounding territory,
and to identify main contributing sources. Sky brightness, naked
eye stellar visibility and telescope limiting magnitude are
produced as 3-dimensional arrays whose axes are the position on
the sky and the atmospheric clarity. We compared our results to
available measurements.

\end{abstract}


%

\section{Introduction}

The change in the light in the night environment due to the
introduction of artificial light is a true pollution, a growing
adverse impact on the night. Pollution means "impairment or
alteration of the purity of the environment" or of its
chemical/physical parameters. This alteration of natural light at
night, called light pollution, can and does impact the environment
and the health of the beings living in it (animals, plants and
man), as shown by hundreds of scientific studies and reports (see
e.g. Rich \& Longcore 2002, Erren \& Piekarski 2002, Cinzano
1994). The growth of the night sky brightness is one of the many
effects of artificial light being wasted in the environment. It is
a serious problem. It endangers not only astronomical observations
but also the perception of the Universe around us (see Crawford
(1991), Kovalewski (1992), McNally (1994), Isobe \& Hirayama
(1998), Cinzano (2000a), Cohen \& Sullivan (2001), Cinzano (2002),
Schwarz (2003) and the International Dark-Sky Association Web
site, www.darksky.org). The starry  sky constitutes mankind's the
only window to the universe beyond the Earth. A fundamental
heritage for the culture, both humanistic and scientific, and an
important part of the our nighttime landscape patrimony is going
to be lost, for those alive today and for our children and their
children. The worldwide growing interest for light pollution and
its effects requires methods for monitoring this situation.

The modelling of the brightness distribution of the night sky at a
site is important to evaluate its suitability for astronomical
observations, to quantify its sky glow, and to recognize
endangered parts of the sky hemisphere. Night sky models are
useful in studying sky glow relationships with atmospheric
conditions and to evaluate future changes in sky glow. The
modelling is also required to disentangle the contribution of
sources, such as individual cities, in order to recognize those
areas producing the strongest impact and to undertake actions to
limit light pollution.

In 1986 Roy Garstang introduced a modelling technique, developed
and refined in the subsequent years (Garstang  1986, 1987, 1988,
1989a, b, 1991a, b, c, 1992, 1993, 2000a), to compute the light
pollution propagation in the atmosphere. He estimated the night
sky brightness at many sites based on the geographical positions,
altitudes and populations of the polluting cities. Cinzano (2000b)
used Garstang models to disentangle the impact of individual
cities constraining free functions with the condition that the sum
of all the contributions with the natural sky brightness fits the
observed sky brightness. However updated population data are not
easily available worldwide, the upward light emission is not
strictly proportional to the population. Some polluting sources,
such as industrial areas and airports, have very low population
density but very high light emission. The U.S. Air Force Defence
Meteorological Satellite Program (DMSP), Operational Linescan
System (OLS) acquires direct observations of nocturnal lighting,
making it possible to map the spatial distribution of nighttime
lights (Sullivan 1989, 1991; Elvidge et al. 1997a, b, c, 2001,
2003a, b; Gallo et al. 2003; Henderson et al. 2003). Most
nighttime OLS observations of urban centers are saturated, making
the data of limited value for modeling purposes. However, Elvidge
et al. (1999) were able to produce a radiance calibrated global
nighttime lights product using OLS data acquired at reduced gain
settings, suitable for the quantitative measurement of the upward
light emission (e.g. Isobe \& Hamamura 2000, Luginbuhl 2001, Osman
et al. 2001) and the evaluation of the artificial sky brightness
produced by it (e.g. Falchi 1998; Falchi \& Cinzano 2000).

Cinzano et al. (2000) presented a method to map the artificial sky
brightness across large territories in a given direction of the
sky by evaluating the upward light emission from DMSP high
resolution radiance calibrated data (Elvidge et al. 1999) and the
propagation of light pollution with Garstang models. A World Atlas
of the artificial night sky brightness at sea level was thus
obtained (Cinzano, Falchi \& Elvidge 2001b). This method was
extended by Cinzano, Falchi \& Elvidge (2001a) to the mapping of
naked eye and telescopic limiting magnitude based on the Schaefer
(1990) and Garstang (2000b) approach and the GTOPO30 elevation
data. We extend and apply their method to the computation of the
distribution of the night sky brightness and the limiting
magnitude over the entire sky at any site for a range of
atmospheric conditions and accounting for mountain screening. In
sec. \ref{maps} we describe the computation on 3-dimensional
arrays whose axes are the position on the sky and the atmospheric
clarity and present our improvements. In sec. \ref{input} we
describe input data. In sec. \ref{dis} we deal with the
disentangling of individual sources. In sec. \ref{appl} we discuss
the application and in sec. \ref{results} we present comparisons
with available measurements. Conclusions are in sec.
\ref{conclusions}.

\section{Computation of the Hypermaps}
\label{maps}

Artificial and natural sky brightness varies depending on the
aerosol content of the atmosphere. The stellar extinction also
vary substantially depending on the aerosol content of the local
atmosphere. This in turn affects the limiting magnitude. So any
map of the sky of a site is a function of the aerosol content for
which it has been computed.

\begin{figure}
\epsfysize=6cm 
\hspace{0.1cm}\epsfbox{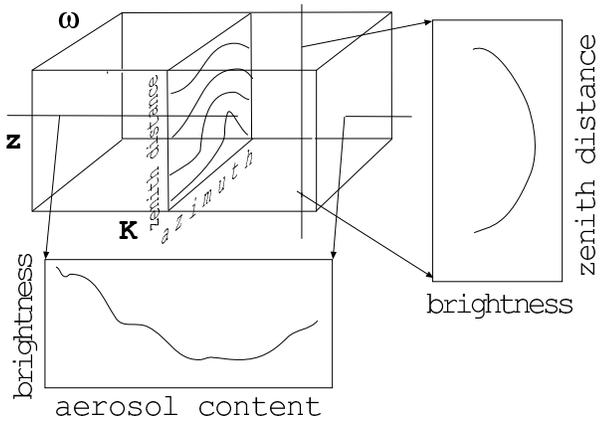} 
\caption[h]{Projections of the hyper-map on different planes.}
\label{f1}
\end{figure}

We refer to a hyper-map as a set of maps of the night sky
brightness for a range of aerosol contents, $b(z,\omega, K)$,
where $z$ is the zenith distance, $\omega$ is the azimuth and $K$
is the aerosol content expressed by the atmospheric clarity
(Garstang 1986, 1989). As fig. \ref{f1} shows, values on planes of
the space of the variables perpendicular to the $K$ axis give maps
of the sky brightness for the given atmospheric clarity, values
along a line parallel to the $K$ axis give the brightness in the
given point of the sky when the atmospheric aerosol content
change, values along lines perpendicular to the $K$ and $\omega$
axis give the sky brightness along an almucantar for the given
atmospheric clarity.

At a site in $(x',\,y')$ the hyper-map is given by:
\begin{equation}
\label{int1} b(z,\omega, K)=\! \! \int_{-\infty }^{+\infty } \! \!
\! \! \int_{-\infty }^{+\infty } \! \! \!\! \! \! e(x,y)
f(x,y,x',y',z,\omega, K)~\mathrm{d}x ~\mathrm{d}y,
\end{equation}
where  $f(x,y,x',y',z,\omega, K)$ is the light pollution
propagation function, i.e. the artificial sky brightness at
$(x',\,y')$ in the direction given by $(z,\omega)$ per unit of
upward light emission $e(x,\,y)$ produced  by the unitary area in
$(x,\,y)$ when atmospheric aerosol content is K. If we divide a
territory in land areas $(h, l)$ with position $(x_{h}, y_{l})$,
the hyper-map can be expressed as a tridimensional array
$b_{i,j,k}$ given by:
\begin{equation}
\label{sum1} b_{i,j,k}=\sum_{h}\sum_{l}  e_{h,l}
f(x_{h},y_{l},x',y',z_{i},\omega_{j},K_{k})\, ,
\end{equation}
where $e_{h,l}$ is the upward flux emitted by the land area
$(h,l)$,  $f(x_{h},y_{l},x',y',z_{i},\omega_{j},K_{k})$ is the
propagation function, $z_{i},\omega_{j},K_{k}$ are an adequate
discretization of the variables $z,\omega, K$ and the sommatories
are extended at all the land areas around the site inside a
distance for which their contributions are non negligible. We
divided the territory in the same land areas covered by pixels of
the satellite data. We obtained the propagation function $f$,
expressed as total flux per unit area of the telescope per unit
solid angle per unit total upward light emission, with models for
the light propagation in the atmosphere based on Garstang models
(Garstang 1986, 1989):
\begin{equation}
\label{gar2}
f \!=\!\! \int^{\infty}_{u_{0}} \! \! \! \! \!\! \! \left( \beta_{\mathrm{m}}(h)
f_{\mathrm{m}}(\varpi)\!+\!\beta_{\mathrm{a}}(h)f_{\mathrm{a}}(\varpi) \right) \\
(1\!+\!D_{\mathrm{S}} ) i(\psi,s) \xi_{1}(u) \mathrm{d}u ,
\end{equation}
where $\beta_{\mathrm{m}}(h)$ $\beta_{\mathrm{a}}(h)$  are
respectively the scattering cross sections  of molecules and
aerosols per unit volume at the altitude $h$, depending on the
distance $u$ along the line of sight of the observer,
$f_{\mathrm{m}}$ and $f_{\mathrm{a}}$ are their normalized angular
scattering functions (see sec.\ref{atmod}), $\varpi$ is the
scattering angle, $\xi_{1}(u)$ is the extinction  of the light
along its path from the scattering volume to the telescope,
$i(\psi,s)$   is the direct illuminance per unit flux produced by
each source on each infinitesimal volume of atmosphere along the
line-of-sight and $(1+D_{\mathrm{S}} )$  is a correction factor
which take into account the illuminance due at light already
scattered once from molecules and aerosols which can be evaluated
as Garstang (1984, 1986), neglecting third and higher order
scattering which can be significant for optical thickness higher
than about 0.5. Geometric relations and formulae accounting for
Earth curvature have been given and discussed by Garstang (1989,
sec. 2.2-2.5, eqs. 4-24). In Garstang's formulae the molecular
scattering cross section per unit volume is
$\beta_{\mathrm{m}}=N_{\mathrm{m}} \sigma_{\mathrm{R}}$.

As Garstang (1989) and differently from Cinzano et al. (2001a) we
take into account the elevation both of the source and of the
site.

Screening by terrain elevation was accounted as described in
Cinzano et al. (2001a). The illuminance per unit flux was set in
eq. (\ref{gar2}) to:
\begin{equation}\label{gar4}
i(\psi,s)=I(\psi) \xi_{2}/ s^{2}\, ,
\end{equation}
where there is no screening by Earth curvature or by terrain
elevation and $i(\psi,s)=0$ elsewhere. Here $I(\psi)$ is the
normalized emission function giving the relative light flux per
unit solid angle emitted by each land area at the zenith distance
$\psi$, $s$ is the distance between the source and the considered
infinitesimal volume of atmosphere and $\xi_{2}$ is the extinction
along the light path, given by Garstang (1989). We check each
point along the line of sight to determine if the source area is
blocked by terrain elevation or not, taking into account Earth
curvature, by determining the position of the foot of the vertical
of the considered point. Then we computed for every land area
crossed by the line connecting this foot and the source area, the
quantity $\cot \psi$ (Cinzano et al. 2001a):
\begin{equation}\label{psi1}
\cot  \psi = \frac{(A+E)-(h+E) \cos (D/E)}{(h+E) \sin (D/E)}\, ,
\end{equation}
where  $A$ is the elevation  of the land area, $D$ is
the distance  of its center from the center of the source area and $E$
is the Earth radius. From it we  determined the screening elevation $h_{\mathrm{s}}$:
\begin{equation}\label{psi2}
h_{\mathrm{s}} = \frac{A+E}{\cos (D^{\star}/E) - \max (-\cot \psi)
\sin (D^{\star }/E)}-E\, ,
\end{equation}
where $D^{\star}$ is the distance between the source area and the
foot of the vertical, and $h_{\mathrm{s}}$ is computed over the
sea level. The illuminance $i$ in eq. (\ref{gar2}) is set to zero
when the elevation of the considered point is lower than the
screening elevation. To speed up the calculation we computed only
once the array, which gives the screening elevation for each point
along the line of sight, for each azimuth of the line of sight and
for each source, and we used it for any computation with different
atmospheric parameters. We considered land areas as point sources
located in their centres except when $i=h$, $j=k$ in which case we
used a four points approximation (Abramowitz \& Stegun 1964). We
assumed that the elevation given by the GTOPO30 be the same
everywhere inside each pixel.

Another array was obtained for the natural sky brightness with the
model introduced by Garstang (1989, sec. 3). The array
$b_{\mathrm{N}\,i,j,k}$ is the sum of (i) the directly transmitted
light $b_{\mathrm{d}}$ which arrives to the observer after
extinction along the line of sight (Garstang 1989, eq. 30), (ii)
the light scattered by molecules by Rayleigh scattering
$b_{\mathrm{r}}$ (Garstang 1989, eq. 37),
 (iii) the light scattered by aerosols $b_{\mathrm{a}}$ (Garstang 1989, eq. 32) :
\begin{equation}\label{nat3}
b_{\mathrm{N}\,i,j,k}=b_{\mathrm{d}\,i,j,k}+ b_{\mathrm{r}\,i,j,k}
+b_{\mathrm{a}\,i,j,k}\, .
\end{equation}
In the computation of the natural sky brightness outside the
scattering and absorbing layers of the atmosphere (Garstang 1989,
eq.29) we assumed as independent variables the brightness of a
layer at infinity, due mainly to integrated star light, diffused
galactic light and zodiacal light, and the brightness of the van
Rhijn layer, due to airglow emission.

The array of the total sky brightness is
$b_{\mathrm{T}\,i,j,k}=b_{\mathrm i,j,k}+b_{\mathrm{N}\,i,j,k}$.
The sky brightness in the chosen photometric band was expressed as
photon radiance (in ph cm$^{-2}$ s$^{-1}$ sr$^{-1}$) or in
magnitudes per arcsec$^{2}$ (Garstang 1989, eqs. 28, 39).

We determined the observer's horizon computing the altitudes below
which the line-of-sight encounter a screen by terrain, like e.g. a
mountain, and set the total brightness to be zero below them. They
are obtained evaluating the elevation $h_\mathrm{T}$ of terrain at
distance $d$ along each azimuth direction and computing the
maximum screening altitude angle $\vartheta$:
\begin{equation}\label{hor1}
 \vartheta=\max \arctan(h_{\mathrm{T}}/d)\, .
\end{equation}

From the array of the total sky brightness in V band we can obtain
a family of other arrays giving the naked-eye star visibility and
the telescopic limiting magnitude. The magnitude over the
atmosphere of a star at the threshold of visibility of an observer
when the brightness of observed background is $b_{\mathrm{obs}}$
in nanolambert and the stimulus size, i.e. the seeing disk
diameter, is $\theta$ in arcmin, has been given by Garstang
(2000b) based on measurements of Blackwell (1946) and Knoll,
Tousey \& Hulburt(1946) and on a threshold criterion of 98 per
cent probability of detection:
\begin{equation}
m_{\mathrm{star}}=-13.98-2.5\log i'_{1} i'_{2} /(i'_{1} +i'_{2})\,
,
\end{equation}
with:
\begin{eqnarray}
\label{g1}
i'_{1}&=& F_{1} c_{1} (1 + k_{1} b^{1/2} )^{2} (1 + \alpha_{1} \theta^{2} + y_{1} b_{\mathrm{obs}}^{z_{1}} \theta^{2})\, , \\
i'_{2}&=& F_{2} c_{2} (1 + k_{2} b^{1/2} )^{2} (1 + \alpha_{2} \theta^{2} + y_{2} b_{\mathrm{obs}}^{z_{2}} \theta^{2})\, , \\
F_{1}&=&F_{\mathrm{a},1}F_{\mathrm{SC},1}F_{\mathrm{cs},1}F_{\mathrm{e},1}F_{\mathrm{s},1}\, , \\
F_{2}&=&F_{\mathrm{a},2}F_{\mathrm{SC},2}F_{\mathrm{cs},2}F_{\mathrm{e},2}F_{\mathrm{s},2}\,
,
\end{eqnarray}
where $i'_{1}$ and $i'_{2}$ are the illuminations produced by the
star, related respectively to the thresholds of scotopic and
photopic vision, and the fraction  is an artifact introduced by
Garstang in order to  put together smoothly the two components
obtaining the best fit with cited measurements. Here
$F_{\mathrm{a}}$ takes into account the ratio between pupil areas
of the observer and the pupil diameter used by the average of the
Knoll, Tousey, Hulburt and Blackwell observers, $F_{\mathrm{SC}}$
takes into account the Stiles-Crawford effect, due to the
decreasing of the efficiency to detect photons with the distance
from the center of the pupil, producing a non linearity in the
increase of sensibility when the eye pupil increase,
$F_{\mathrm{cs}}$ allows for the difference in color between the
laboratory sources used in determining the relationships between
$i$ and $b$ and the observed star, $F_{\mathrm{e}}$ allows for
star light extinction in the terrestrial atmosphere because star
magnitudes are given {\it outside the atmosphere},
$F_{\mathrm{s}}$ allows for the acuity of any particular observer,
defined so that $F_{\mathrm{s}}< 1$ implies an eye sensitivity
higher than average due possibly to above average retinal
sensitivity, scientific experience or an above average eye pupil
size. Formulae have been given by Schaefer (1990) and Garstang
(2000b) and applied by Cinzano et al. (2001a, eqs. 28-31) to which
we refer the reader. The constants $c$, $k$, $\alpha$, $y$, $z$ in
eq. (\ref{g1}) are given by Garstang (2000b). The perceived
background $b_{\mathrm{obs}}$ is related to the total sky
brightness under the atmosphere in $V$ band given by our
hyper-maps, converted from photon radiance to nanolambert
(Garstang 2000b):
\begin{equation}
b_{\mathrm{obs}}=b_{\mathrm{T}}/(F_{\mathrm{a}}F_{\mathrm{SC}}F_{\mathrm{cb}})
\, ,
\end{equation}
where $F_{\mathrm{cb}}$ allow for the difference in color between
the laboratory sources and the  night sky background, and
$F_{\mathrm{a}}, F_{\mathrm{SC}}$ were already described. As a
result we obtain the array $m_{i,j,k}$ of the visual limiting
magnitude. The array of the telescopic limiting magnitudes can be
calculated for the chosen instrumental setup in a similar way (see
the cited authors).

\section{Input data} \label{input}

We summarize here the required input data, which has been already
described and discussed by Cinzano et al. (2000, 2001a). We refer
the reader to their paper for details. We extended the input data
to other continents in the same way.

\subsection{Upward light emission data}
\label{upw}

To compute the illuminance per unit flux $i$ in eq. \ref{gar4} we
need the relative intensity $I(x,y,\psi,\chi)$ emitted by every
land area in $(x,y)$ at azimuth $\chi$ and zenith distance $\psi$,
i.e. the normalized emission function obtained measuring the
relative emitted flux per unit solid angle per unit area in the
direction $\psi$ and normalizing its integral to unity. If the
land areas contain many light installations randomly distributed
in type and orientation, we can assume this function axysimmetric
$I(x,y,\psi)$. The corresponding absolute intensity is:
\begin{equation}
I'(x,y,\psi)=e(x,y)\times I(x,y,\psi)\, ,
\end{equation}
where $e(x,y)$ is the total upward flux obtained from radiance
calibrated data (Cinzano 2001a, eq. 35).

We obtained the upward flux $e(x,y)$ on a 30''$\times$30'' pixel
size grid from the Operational Linescan System (OLS) carried by
the DMSP satellites after a special requests to the U.S. Air Force
made by the U.S. Department of Commerce, NOAA National Geophysical
Data Center (NGDC), which serves as the archive for the DMSP and
develops night time lights processing algorithms and products. OLS
is an oscillating scan radiometer with low-light visible and
thermal infrared (TIR) high-resolution imaging capabilities
(Lieske 1981). The OLS Photo Multiplier Tube (PMT) detector has a
broad spectral response covering the range for primary emissions
from the most widely used lamps for external lighting. The primary
reduction steps were (Cinzano et al. 2000, 2001a; Elvidge et al.\
1999):
\begin{enumerate}
\item 1) acquisition of special OLS-PMT data at a number of reduced
gain settings to avoid saturation on major urban centres and, in
the same time, overcome PMT dynamic range limitation.  On board
algorithms which adjust the visible band gain  were disabled.
\item establishment of a reference grid with finer spatial resolution
than the input imagery;
\item identification of the cloud free section of each orbit based on OLS-TIR data;
\item identification of lights, removal of noise and solar glare,
cleaning of defective scan lines;
\item  projection of the lights from cloud-free areas from each orbit
into the reference grid;
\item  calibration to radiance units using preflight calibration of
digital numbers for given input telescope illuminance and  gain settings in
simulated space conditions;
\item  tallying of the total number of light detections in each grid cell and
calculation of the average radiance value;
\item  filtering images based on frequency of detection to remove ephemeral events;
\item  transformation  into latitude/longitude projection with 30''$\times$30'' pixel size;
\item  Lucy-Richardson deconvolution to improve predictions for sites near sources
(when possible this should be more properly done before step 7).
\item  Determination of the upward light intensity accounting for the estimated
atmospherical extinction in the light path from ground to the
satellite, the assumed average spectrum of  night-time lighting
(Cinzano et al. 2000a, eqs. 28-30) and the surface of the land
areas. \end{enumerate} We can obtain $I(x,y,\psi)$ from the
radiance measured in a set of individual orbit satellite images
where the land area in $(x,y)$ is seen at different angles $\psi$
which are related to the distance $D$ from the satellite nadir
(Cinzano et al. 2000, eq. 17, 18). The emitted flux per solid
angle per unit area in the direction $\psi$ is obtained from the
measured radiance dividing by the extinction coefficient
$\xi_{3}(\psi)$ computed for curved-Earth (Cinzano et al. 2000,
eq. 19). A study to obtain $I(x,y,\phi)$ in this way for every
land area from DMSP-OLS individual orbit data is in progress
(Cinzano, Falchi, Elvidge, in prep.). To be simple we assumed here
that all land areas have in average the same normalized emission
function, given by the parametric representation of Garstang
\cite{g86} in eq. (15) of Cinzano et al. (2000), which has been
tested by studying in a single orbit satellite image the relation
between the upward flux per unit solid angle per inhabitant of a
large number of cities and their distance from the satellite nadir
(Cinzano et al. 2000) and with many comparisons between model
predictions and measurements by Garstang and by Cinzano (2000b).
Likely it cannot be applied in areas where effective laws against
light pollution are enforced or with unusual lighting habits.

\subsection{Elevation data}

As input elevation data we used GTOPO30, a global digital
elevation model by the U.S. Geological Survey's EROS Data Center
(Gesch et al. 1999). This global data set covers the full extent
of latitude and longitude with an horizontal grid spacing of 30''
as our composite satellite image. The vertical units represent
elevation in meters above mean sea level which ranges from -407 to
8,752 meters. We reassigned a value of zero to ocean areas, masked
as "no data" with a value of -9999, and to altitudes under sea
level.

\subsection{Atmospheric data}
\label{atmod}

In order to evaluate scattering and extinction we need a set of
functions giving, for each triplet of longitude, latitude and
elevation $(x, y, h)$, the molecular and aerosol cross scattering
coefficients per units volume of atmosphere
$\beta_{\mathrm{m}}(x,y,h)$ and $\beta_{\mathrm{a}}(x,y,h)$, and
the aerosol angular scattering function $f_{\mathrm{a}}(\omega, x,
y, h)$. The molecular angular scattering function
$f_{\mathrm{m}}(\omega)$ is known because it is Rayleigh
scattering. The atmospheric data need to refer  at a {\it typical}
clean night in the chosen time of the year and to include
information on denser aerosol layers, volcanic dust and Ozone
layer.

To be simple we  applied here the standard atmospheric model
already adopted by Garstang (1986, 1989) and Cinzano et al. (2000,
2001a), neglecting geographical gradients and local
particularities. It assumes:
\begin{enumerate}
\item the molecular atmosphere in hydrostatic equilibrium under the gravitational
force as \cite{g86}.
\item  the atmospheric haze aerosols number density decreasing exponentially as
\cite{g86}
\item  a neglegible presence of sporadic denser aerosol layers, volcanic dust and
Ozone layer (studied by Garstang 1991a, c).
\item  the normalized angular scattering function for atmospheric haze aerosols
given in \cite{g91a}.
\item  the aerosol content given by an atmospheric clarity
parameter which measures the relative importance of aerosol and
molecules for scattering light.
\end{enumerate}
The Garstang atmospheric clarity parameter $K$ measures the
relative importance of aerosol and molecules for scattering light
in $V$ band at ground level (Garstang 1996):
\begin{equation}\label{kappa}
K=\frac{\beta_{\mathrm{a},H} }{\beta_{\mathrm{m,0}}  11.11
\mathrm{e}^{-cH}}\,  ,
\end{equation}
where $H$ is the altitude of the ground level over sea level and
$c$ is the inverse scale height of molecules. It assumes that
there is only one ground level where all the polluting sources
lie. To be more self-consistent when there are many cities at
different elevations over sea level, we introduced an atmospheric
clarity parameter $K'$ defined at sea level:
\begin{equation}\label{kappa1}
K'=\frac{\beta_{\mathrm{a,0}} }{\beta_{\mathrm{m,0}}11.11}  \,  ,
\end{equation}
so that at ground level of each city $K=K'\mathrm{e}^{(c-a)H}$,
where $a$ is the inverse scale height of aerosols. We can
associate the atmospheric clarity K with vertical extinction (e.g.
Garstang 1991, eq. 6) and with other observable quantities like
the horizontal visibility (Garstang 1989, eq. 38), the optical
thickness $\tau$ (Garstang 1986, eq. 22) and the Linke turbidity
factor for total solar radiation (Garstang 1988). Extinction along
light paths for this atmospheric model was given by Garstang
(1989, eqs. 18-22).

\subsection{Natural night sky brightness data}
\label{natural}

The brightness $b_{\mathrm{S}\,i,j}$, due to integrated star
light, diffused galactic light and zodiacal light, depends on the
observed area of the sky and on the time. This dependence  on the
position of the sky is important when sky maps are made to
quantify the visibility of  astronomical phenomena, otherwise we
can assume $b_{\mathrm{S}\,i,j}$ constant and given by its average
value in the considered site. The brightness of the Van Rhijn
layer, $b_{\mathrm{VR}}$, depends on some factors like the
geographical position, the solar activity in the previous day, and
the time after twilight. We referred our predictions to some hours
after the twilight, when the night brightness decay at a constant
value (Walker 1988, but see also Patat 2003a), and to minimum
solar activity. If requested, the solar activity can be roughly
accounted as Cinzano et al (2001a) or, more accurately, based on
the correlation with the 10.7 cm solar radio flux (Walker 1988,
Krisciunas 1999). The dependence of $b_{\mathrm{VR}}$ by the
geographical position suggests to study the natural sky brightness
in the nearest unpolluted site, which can be located in the world
atlas of artificial sky brightness (Cinzano et al. 2001b), in
order to obtain $b_{\mathrm{S}\,i,j}$ and $b_{\mathrm{VR}}$. When
only one or few measurements were available we assumed as Garstang
(1989) $b_{\mathrm{S}\,i,j}=0.4 b_{0}$ and $b_{\mathrm{VR}}=0.6
b_{0}$ and determined $b_{0}$.

\section{Disentangling individual contributions} \label{dis}

We can make some analysis on the contributions from each
30''$\times$30'' land area which enter in the summatory of eq.(2).
First we can make hypermaps of sky brightness produced by
individual land areas and compare them. Moreover, chosen an array
cell of index $(i,j,k)$ we can obtain a geographic map showing the
contribution $b_{i,j,k}(x_{h},y_{l})$ produced by each land area
in $(x_{h},y_{l})$, searching for main polluting sources and
making some statistic on their geographic distribution:
\begin{equation}
b_{i,j,k}(x_{h},y_{l})=e_{h,l}
f(x_{h},y_{l},x',y',z_{i},\omega_{j},K_{k})\, .
\end{equation}
We can obtain hypermaps of sky brightness produced by each city or
territory identifying pixels belonging to each city or territory
of a given list and summing their contributions:
\begin{equation}
b_{i,j,k}(n)=\mathop{\sum\nolimits_{h,l}}\limits_{\mathrm{ n\!\!
-\! th \,city}} e_{h,l}
f(x_{h},y_{l},x',y',z_{i},\omega_{j},K_{k})\, .
\end{equation}
Their comparison is helpful e.g. to understand if larger
contributions come from few main cities or from many small towns,
even in relation with atmospheric conditions. The fraction of sky
brightness produced in a given array cell $(i,j,k)$ by the sources
inside a circular area of radius $d$ can be obtained summing all
contributions of land areas inside the distance $d$ from the site
and dividing by the sum of all contributions:
\begin{equation}
b^{\star}(d)= \frac{1}{b_{i,j,k}}\! \! \! \! \!
\mathop{\sum\nolimits_{h,l}}\limits_{\scriptstyle (x_{h}\!
-x')^{\!2}+ \hfill \atop \scriptstyle \;   (y_{l}\! -y')^{\!2} \!
\leq d^{2} \hfill } \! \!\! \! \! \! e_{h,l}
f(x_{h},y_{l},x',y',z_{i},\omega_{j},K_{k})\, .
\end{equation}
This is useful e.g. to evaluate the effectiveness of protection
areas (Cinzano 2000c).

\section{Application}\label{appl}

The software package {\sc lpskymap}, written in Fortran-77,
calculates the artificial night sky brightness, the total night
sky brightness and the star visibility (limiting magnitude) over
the entire sky at any site in the World. The availability of
OLS-DMSP fixed gain data on a yearly or sub-yearly timescale will
allow a fine time resolution.

Results are arrays of the artificial night sky brightness, the
total night sky brightness, the visual limiting magnitude and the
loss of visual limiting magnitude. Each hyper-map array is
composed by a series of 19$\times$37 pixel images in cartesian
coordinates, one for each aerosol content $K$, spline interpolated
over 91$\times$181 pixels in cartesian coordinates or projected in
721$\times$721 pixels in polar coordinates. Images go from 0 to
360 degrees in azimuth, starting from East (in order to avoid to
place the meridian at borders) toward South, and from horizon to
zenith in altitude. They are saved in 16-bit standard {\sc fits}
format with {\sc fitsio} Fortran-77 routines developed by HEASARC
at the NASA/GSFC. ASCII data tables are also provided. The night
sky brightness in the chosen photometric band is given as photon
radiance in ph s$^{-2}$ m$^{-2}$ sr$^{-1}$ or as astronomical
brightness in mag arcsec$^{-2}$. Brightness in V band can be also
expressed as luminance in $\mu$cd m$^{-2}$, using Garstang's
conversion (Garstang 2002; Cinzano 2004). From the hyper-map
arrays we can obtain:
\begin{enumerate}
\item sections perpendicular to the $K$ axis: $b(z,\omega,K\!=\!K_{0})$.
They are the maps of the sky brightness or limiting magnitude for
a given aerosol content and they correspond to each individual
image of the series.
\item secants parallel to the $K$ axis:
$b(z\!=\!z_{0},\omega\!=\!\omega_{0},K)$. They provide the
brightness or the limiting magnitude in a given point of the sky
as the aerosol content changes.
\item secants perpendicular to the $K$ and $\omega$ axis:
$b(z,\omega\!=\!\omega_{0},K\!=\!K_{0})$. They give the brightness
or the limiting magnitude along an almucantar, e.g. the meridian,
for a given aerosol content.
\end{enumerate}

The arrays computation steps are:
\begin{enumerate}
\item An input file is prepared with the geographical position and elevation
of the site, the names of input DEM and lights frames and the
position of their upper left corner.
\item The array, i.e. the series of images, of the artificial night
sky brightness is computed with the program {\sc lpskymap} for a
given range and step of the aerosol content, accounting for Earth
curvature and elevation but not for screening. The radius of the
contributing area can be 250 km for sites in urbanized areas or
350 km for dark sites.
\item Subimages with DEM and lights data have been cropped from
the original large scale frames with the program {\sc makefrac}.
We use {\sc fits} or {\sc raw} images 701x701 px in size to limit
the requirements of RAM memory during screening computation. They
are checked for relative mismatches which can be corrected with
the program {\sc makeshift}.
\item The screening angles for each direction of observation and for each
area inside a given radius from the site are computed with the
program {\sc makescreen}. We limited the radius to 200 km to avoid
too long computation time. The program writes the screening data
of each site in 106 files for a total size of 20GB uncompressed.
It also calculates the horizon line as seen by the site. DEM
pixels very near to the site are divided in 11x11 sub pixels
evaluated separately.
\item An array containing the screened brightness is computed with the
program {\sc lpskyscreen} when there are reasons to believe that
screening is not negligible.
\item The images of the screened brightness array are
subtracted from the correspondent images of the sky brightness
array, after properly rescaling, in order to obtain the array of
the night sky brightness corrected for mountain screening.
\item The array is calibrated with the program {\sc lpskycal} based on
pre-flight calibration at 1996-1997, or on Cinzano et al. (2001b)
calibration at 1998-1999 made with Earth-based measurements, or on
observations taken at the same site. Measurements of Cinzano et
al. (2001a) fitted predictions based on the pre-flight calibration
with $\sigma\leq$0.35 mag arcsec$^{-2}$ and a shift $\Delta
m$=-0.28 mag arcsec$^{-2}$, likely mainly due to the growth of
light pollution in the period between the observations and the
satellite data acquisitions. The program adds the natural sky
brightness, producing a series of calibrated maps of the total
night sky brightness, interpolated or not, and the limiting
magnitude. It also adds the horizon line. It does not account for
the refraction of light by the atmosphere which could increase the
brightness near the horizon toward very far cities.
\item Maps in polar coordinates are obtained with the program
{\sc lpskypolar}. East is up, North at right.
\item Maps are analyzed with {\sc ftools} developed by HEASARC
at the NASA/GSFC.
\item Comparison with observations is made with the program
{\sc lpskycompare}. Measurements should be "under the atmosphere".
Statistical analysis is made with the software {\sc mathematica}
of Wolfram Research.
\end{enumerate}
A number of utility programs completes the package. The
computation time depends on the geographical behavior of the site,
like the quantity of dark pixels, the quantity of nonzero
elevation pixels, etc. As an example, the computation of one
element of the array (i.e. a single map for a given atmospheric
content) for Sunrise Rock on a workstation with Xeon processor
running at 1700 MHz required about 35 hours for {\sc lpskymap}, 10
hours for {\sc makescreen} and 6 hours for {\sc lpskyscreen}.
However, the computation with {\sc lpskymap} for the site in Padua
required 80 hours, even if restricted inside a radius of 250 km,
whereas the same computation for Serra La Nave required 18 hours
only.

\section{Results} \label{results}

In this section we present a sample of results which can be
obtained with our method and some comparisons with available
measurements. Specific studies are reserved for forthcoming
papers.

NGDC request for low and medium gain DMSP-OLS data used in this
work was granted from U.S. Air Force for the darkest nights of
lunar cycles in March 1996 and January-February 1997. More recent
data sets taken in the period 1999-2003 are already at our
disposal, but they are still under reduction and, before we are
able to use them, we need to solve a number of problems in the
analysis process (Cinzano, Falchi \& Elvidge, in prep.).
Pre-flight calibration of upward flux refers to 1996-1997, to the
average lighting spectra of Cinzano et al. (2000) and to an
average vertical extinction in {\it V} band at imaging time
assumed to be $\Delta m = 0.33$ mag. All results are computed for
minimum solar activity and refers to some hours after twilight. We
tuned the parameter $b_{0}$ to fit the zenith natural sky
brightness for clean atmosphere measured by Cinzano et al. (2001a)
at Isola del Giglio, Italy, $V=21.74\pm0.06$ mag arcsec$^{-2}$ in
{\it V} band for average solar activity and 200 m altitude over
sea level. It agrees well with the average natural night sky
brightness of 21.6 mag arcsec$^{-2}$ measured by Patat (2003a) at
ESO-Paranal. In facts, the sky become darker going to lower
elevation over sea level due to larger extinction, even if this
phenomena is limited by the increase of the light scattered from
aerosols and molecules along the line of sight (Garstang 1989).
Patat (2003) reported a large contribution from zodiacal light,
about 0.18 mag arcsec$^{-2}$, which justifies the fact that he
finds the sky slightly more luminous than expected. The algorithm
of Patat (2003b) applied to VLT images excludes almost completely
the stellar component whereas Cinzano et al. (2001a) excluded only
stars fainter than 18th mag, but the expected difference is only
$\approx$0.03 mag arcsec$^{-2}$. The "visual" natural night sky
brightness should be obtained from the measured one adding the
average stellar background produced by stars with magnitude
$\geq$7 missed by the instrument or the analysis (Cinzano \&
Falchi 2004). This contribute is about -0.26 mag arcsec$^{-2}$
when stars down to magnitude 24 are missed. In our brightness
predictions we did not correct the natural night sky brightness to
the visual value.

\begin{figure}
\epsfysize=6cm 
\hspace{0.1cm}\epsfbox{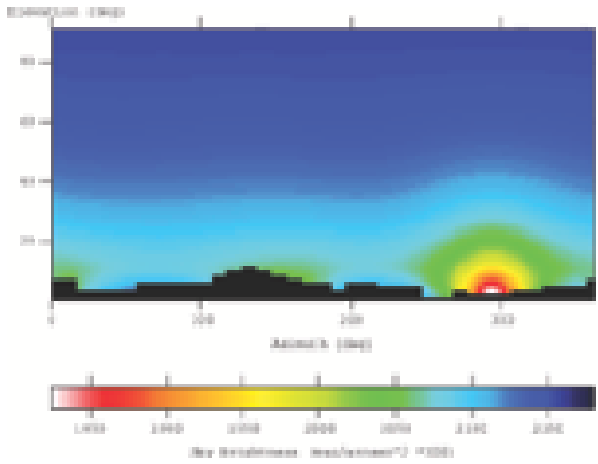} 
\caption[h]{Night sky brightness at Sunrise Rock, USA for
atmospheric clarity $K'$=0.5.} \label{f2}
\end{figure}

\begin{figure}
\epsfysize=3.4cm 
\hspace{0.7cm}\epsfbox{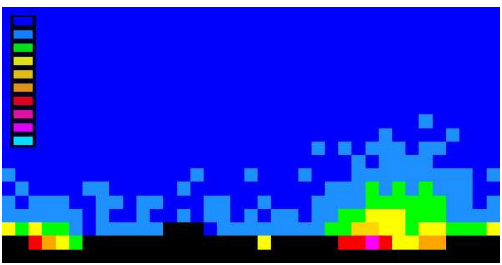} 
\caption[h]{Brightness screened by mountains at Sunrise Rock, USA
for atmospheric clarity $K'$=0.5. Each level from blue to violet
is 0.01 mag arcsec$^{-2}$.} \label{f2b}
\end{figure}

\begin{figure}
\epsfysize=6cm 
\hspace{0.1cm}\epsfbox{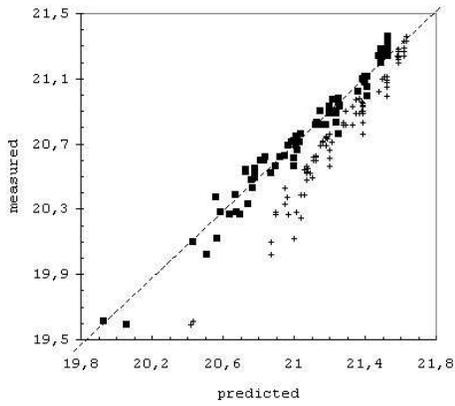} 
\caption[h]{Comparison between predictions and V band measurements
at Sunrise Rock for atmospheric clarities $K'$=0.5 (squares) and
$K'$=3 (crosses). Units are mag arcsec$^{-2}$.} \label{f3}
\end{figure}

Fig. \ref{f2} shows the night sky brightness at Sunrise Rock, a
site located in Mojave National Preserve, California, USA (long W
115$\degr$ 33' 6.4'', lat N 35$\degr$ 18' 57.7'') at 1534 m over
sea level. This site is mainly polluted by the lights of Las
Vegas, about 100 km away. Azimuth goes from 0 to 360 degrees,
starting from East toward South. Fig. \ref{f2b} shows the night
sky brightness screened by mountains, which amounts to few
hundredth of magnitude. Fig. \ref{f3} shows a comparison between
predictions for atmospheric clarities $K'$=0.5 (squares) or $K'$=3
(crosses) and {\it V} band measurements taken on 2003 May 8 at
05.34-06.00 UT (Duriscoe et al. 2004) with vertical extinction
k$_{V}$=0.18 mag. The agreement is excellent after an uniform
scaling of about -0.3 mag arcsec$^{-2}$. It suggests an increase
of light pollution from 1997 to 2003 of $\approx$5 per cent per
year, slightly less than the average yearly growth $\approx$6 per
cent estimated by Cinzano (2003). A comparison with a data set
taken on 2003 September 22 at 06.27-06.58 UT with k$_{V}$=0.26 mag
gives similar results.

\begin{figure}
\epsfysize=6cm 
\hspace{0.1cm}\epsfbox{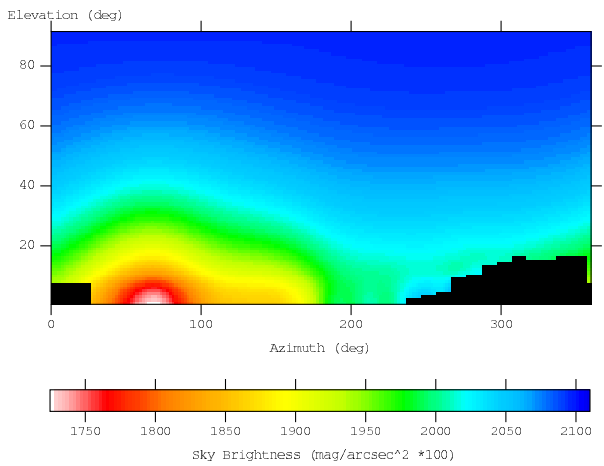} 
\caption[h]{Night sky brightness at Serra la Nave Observatory,
Italy for atmospheric clarity $K'$=1.} \label{f4}
\end{figure}

\begin{figure}
\epsfysize=6cm 
\hspace{0.1cm}\epsfbox{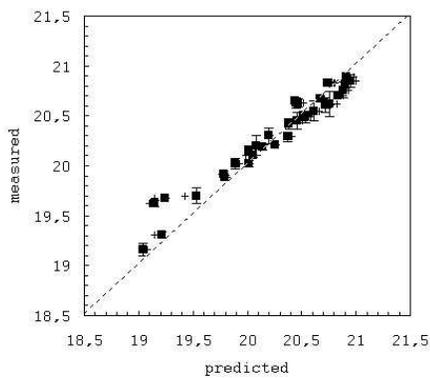} 
\caption[h]{Comparison between predictions and V band measurements
at Serra la Nave Observatory for atmospheric clarities $K'$=1
(squares) and $K'$=2 (crosses). Units are mag arcsec$^{-2}$.}
\label{f5}
\end{figure}

Fig. \ref{f4} shows the night sky brightness at Serra la Nave
Observatory (long E 14$\degr$ 58' 24'', lat N 37$\degr$ 41' 30'')
at 1734 m over sea level on the Mt. Etna volcano, Italy. This site
is situated at few kilometers from a densely populated area with
 $\sim$1.8 10$^{6}$ inhabitants, which includes the cities of
Catania (23 km) and Messina (75 km). Fig. \ref{f5} shows a
comparison between predictions for atmospheric clarities $K'$=1
(squares) and $K'$=2 (crosses) with {\it V} band measurements
taken on 1998 February 22-23 at 18.00-20.00 UT with vertical
extinction k$_{V}$=0.26 mag (Catanzaro \& Catalano 2000; see fig.
2). The agreement is good. The fit is slightly better for the
model with $K'$=1, corresponding to a vertical extinction of
k$_{V}$=0.17 mag, which is smaller than the measured one. However
the vertical extinction at this site could be locally determined
by the volcanic dust (Catanzaro, priv. comm.) whereas $K'$ depends
on the average aerosol content of the entire area with 250 km
radius, so they do not need to match.

The effect of an increase of the aerosol content depends on the
distribution of sources around the site. In general it decreases
the zenith brightness when the distance of the main sources is
larger than few kilometers, decreases the brightness at low
elevation in direction of far sources, increases the brightness at
very low elevation in direction of sources at small or average
distance. This could explain the different behavior of the sky
brightness with the aerosol content in these two sites.

\begin{figure}
\epsfysize=5cm 
\hspace{0.1cm}\epsfbox{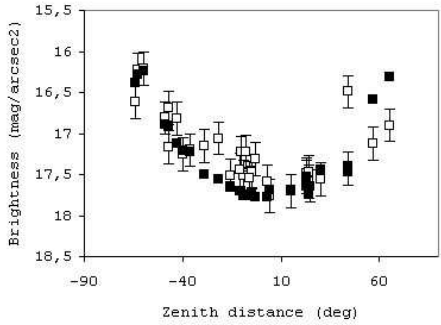} 
\caption[h]{Brightness - zenith distance relation measured at G.
Ruggeri Observatory, Italy (open symbols) and predictions for the
same viewing directions (filled symbols) for atmospheric clarity
$K'$=3 versus the zenith distance. Positive elevations collect
measurements with zenith distances less than $\pm$90$\degr$ from
the direction of the city centre.} \label{f6}
\end{figure}

\begin{figure}
\epsfysize=6cm 
\hspace{0.1cm}\epsfbox{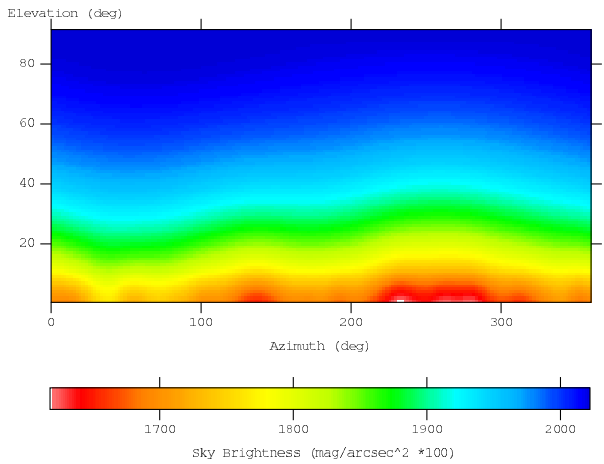} 
\caption[h]{Contribution to the artificial night sky brightness at
Padua from the sources outside Padua for atmospheric clarity
$K'$=1.} \label{f7}
\end{figure}

\begin{figure}
\epsfysize=4.5cm 
\hspace{1cm}\epsfbox{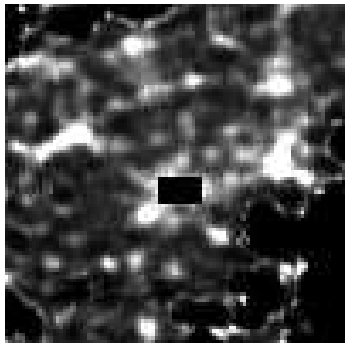} 
\caption[h]{Distribution of lights in the plain surrounding Padua
from OLS-DMSP satellite data. Dark section is the neglected area
in the prediction of Fig. \ref{f7}. The region shown is 50' square
in geographic latitude/longitude projection (approximately 65
$\times$ 93 km).} \label{f8}
\end{figure}

 Fig. \ref{f6} shows the night sky brightness versus the zenith
 distance at G. Ruggeri Observatory, Padova,
Italy (long E 11$\degr$ 53' 20'', lat N 45$\degr$ 25' 10''). This
site is located inside a city of 8 $10^5$ inhabitants in a plain
with more than 4 $10^6$ inhabitants. Positive zenith distances
collect measurements with azimuth inside $\pm$90$\degr$ from the
direction of the city centre. Open symbols are the {\it V} band
measurements taken on 1998 March 26 at 20.00-23.30 UT, with
k$_{V}$=0.48 mag (Favero et al 2000). Filled symbols are
predictions in the same directions for atmospheric clarity $K'$=3,
corresponding to k$_{V}$=0.65 mag. For smaller values of $K'$ the
brightness is underestimated by a constant value. This is likely
due to the fact that our model cannot accurately account for the
scattered light coming from lighting installations inside few
hundreds of meters from the site because pixel sizes are of the
order of 1 km. We used for this prediction the calibration made
for 1998-1999 by Cinzano et al. (2001a). For an atmospheric
clarity $K \geq 2.2$, i.e. for an optical depth $\tau\geq 0.5$,
the double scattering approximation could be not fully adequate
(Garstang 1989; Cinzano et al. 2000). Fig. \ref{f7} shows the
contribution to the artificial night sky brightness produced in
the same site from the sources outside Padua for atmospheric
clarity $K=$1.9 (k$_{V}$=0.48 mag). The area neglected in the
prediction is shown in Fig. \ref{f8} together with the
distribution of lights in the Padana Plain surrounding Padua from
OLS-DMSP satellite data.

\begin{figure}
\epsfysize=6.5cm 
\hspace{0.5cm}\epsfbox{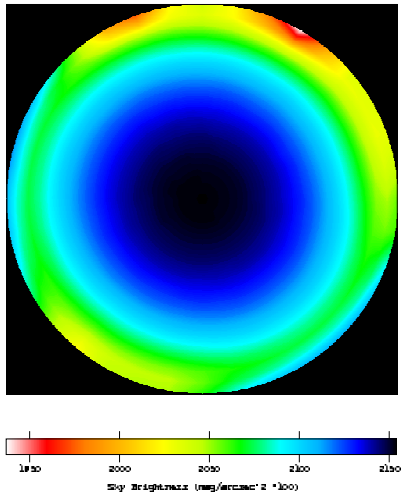} 
\caption[h]{Night sky brightness in V band at Mt. Graham
Observatory, USA in polar coordinates for atmospheric clarity
$K'=0.5$. The figure is plotted with East at bottom, North at
left.} \label{f9}
\end{figure}

Fig. \ref{f9} shows in polar coordinates the total night sky
brightness in V band at Mt. Graham Observatory, USA (long W
109$\degr$ 53' 31'', lat N 32$\degr$ 42' 5'', 3191m o.s.l.) for
atmospheric clarity $K'=0.5$. It can be compared with the image
available at the web address http://mgpc3.as.arizona.edu/images/
Night\%20Sky\%20large.jpg or with fig. 8 of Garstang (1989), which
shows only the artificial brightness.

\begin{figure}
\epsfysize=6cm 
\hspace{0.1cm}\epsfbox{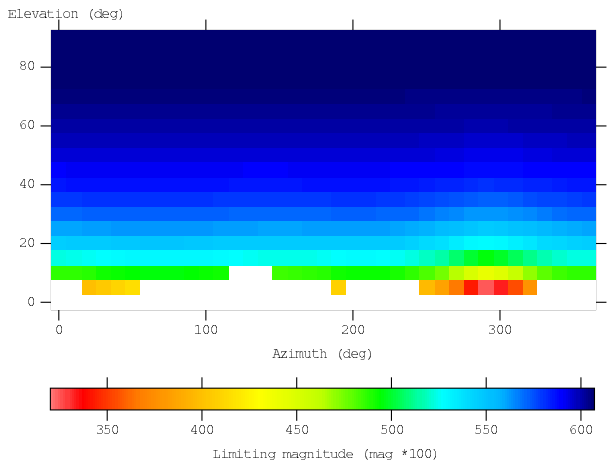} 
\caption[h]{Naked eye limiting magnitude at Sunrise Rock, USA for
atmospheric clarity $K'=0.5$ and 98 per cent detection
probability.} \label{f10}
\end{figure}

Fig. \ref{f10} shows the naked eye limiting magnitude at Sunrise
Rock. Limiting magnitude is computed for observers of average
experience and capability $F_{s}=1$, aged 40 years, 98 per cent
detection probability (faintest star that the observer sees {\it
surely} and not the faintest {\it suspected} star) and star color
index $B-V$ = 0.7 mag. Experienced amateur astronomers are more
trained in naked-eye observation than unexperienced peoples and
can consider detected a star at a much smaller detection
probability so that their limiting magnitude can be more than one
magnitude larger (Schaefer 1990). See the discussion in Cinzano et
al. (2001a).

\begin{figure}
\epsfysize=6cm 
\hspace{0.1cm}\epsfbox{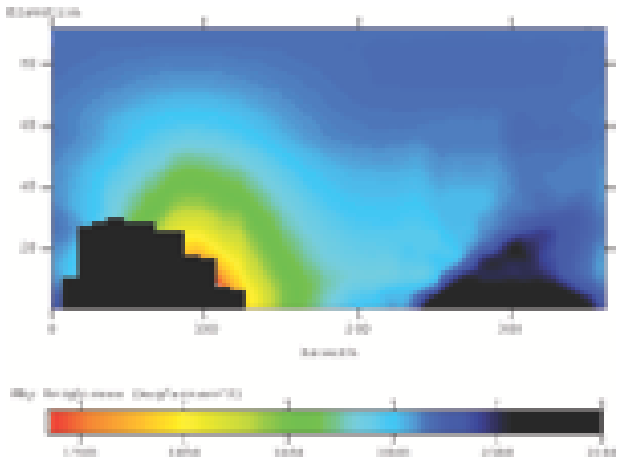} 
\caption[h]{A city screened by a large mountain (left), off-set in
respect to the line observer-source, projects an asymmetric
Schaefer's umbra on the sky (right). Brightness scale is
arbitrary. } \label{f11}
\end{figure}

We checked the effects of the mountain screening trying to
reproduce the umbrae on the sky modelled by Schaefer (1988) and
due to the screening produced by the Mauna Kea on the light of the
rising sun backscattered to the observer. Fig. \ref{f11} shows the
analogous of the Schaefer's umbrae produced by a source of light
pollution instead of the Sun. A city screened by a large conic
mountain (left) projects an umbra over the horizon (right). When
the mountain is off-set in respect to the line observer-source, a
non symmetric penumbra appears. Here the penumbra is at higher
altitudes than in the Wynn-Williams' photo (Schaefer 1988, fig.1)
likely because the observer is at lower elevation. Further
examples of umbrae and baffles are shown by Cinzano \& Elvidge
(2003a fig.1 and fig.3, 2003b).

\section{Conclusions} \label{conclusions}

We extended the seminal works of Garstang by applying his models
to upward flux data from DMSP satellites and to GTOPO30 digital
elevation models, and by accounting for mountain screening. The
presented method allows one to monitor the artificial sky
brightness and visual or telescopic limiting magnitudes at
astronomical sites or in any other site in the World.

This study provides fundamental information in evaluating
observing sites suitable for astronomical observations, to
quantify  sky glow, to recognize endangered parts of the sky
hemisphere when measurements are not readily available or easy
feasible, and to quantify the ability of the resident population
to perceive the Universe they live in. The method enables to study
the relationship of night sky brightness with aerosol content and
to evaluate its changes with time. The method also allows one to
analyze the adverse impacts on a site from the surrounding
territories, making it possible to disentangle individual
contributions in order to recognize those that are producing the
stronger impact and hence to undertake actions to limit light
pollution (the use of fully shielded fixtures, limitation of the
downward flux wasted outside the lighted surface, use of lamps
with reduced scotopic emission, flux reduction whenever possible,
no lighting where not necessary, restraining of lighting growth
rates or lighting density, etc). We also present some tests of the
method. The effects of light pollution on the night sky are easily
evident in the maps in the text.

Important refinements needs to be done in the future years: i) it
may be possible to derive the angular distribution of light
emissions from major sources of nighttime lighting from OLS or
future satellite data (Cinzano, Falchi, Elvidge in prep.). This
will improve accuracy of the modelling, in particular where laws
against light pollution are enforced; ii) a global Atlas of the
growth rates of light pollution and zenith night sky brightness
from satellite data (Cinzano, Falchi, Elvidge in prep.) will make
it possible to predict the evolution of the night sky situation at
sites; iii) a worldwide atmospheric data set giving the
atmospheric conditions in any land area for the same nights of
satellite measurements or for a typical local clear night will
allow to replace the standard atmosphere with the true atmosphere
or the typical local atmosphere; iv) the availability of spectra
of the light emission of each land area from satellite will allow
a more precise conversion of OLS data to astronomical
photometrical bands and an accurate modelling of the colors of the
night sky; v) a large number of accurate measurements of night sky
brightness and visual limiting magnitude including the evaluation
of the atmospheric content, from  e.g. the vertical extinction,
will allow to better constrain predictions allowing improvements
of the modelling technique. The International Dark-Sky
Association, the organization which takes care of the battle
against light pollution and the protection of the night sky is
making a large worldwide effort to collect accurate measurements
of both night sky brightness and stellar extinction (e.g. Cinzano
\& Falchi 2004). They constitute a fundamental component of the
monitoring of the night sky situation in the World.

\section*{Acknowledgments}
We are indebted to Roy Garstang of JILA-University of Colorado for
his friendly kindness in reading this paper, for his helpful
suggestions and for interesting discussions. We acknowledge the
EROS Data Center, Sioux Falls, USA for kindly providing us their
GTOPO30 digital elevation model and the National Park Service
Night Sky Team, Death Valley, USA, together with the authors Dan
Duriscoe, Chad Moore, and Christian Luginbuhl, for kindly
providing us some of their night sky brightness data. This work
has been supported by the Italian Space Agency contract
I/R/160/02. The application to Padua is part of a research project
supported by the University of Padua CPDG023488.

\onecolumn

\twocolumn


{}

\label{lastpage}

\bsp

\end{document}